\documentclass[prc,twocolumn,floatfix,showpacs,superscriptaddress]{revtex4}
\usepackage{amssymb}
\usepackage[centertags]{amsmath}
\usepackage{txfonts}
\usepackage{bm}
\usepackage{epsfig}
\usepackage{graphicx,graphics}
\usepackage{float}

\begin{document}
\title{ Hadronization effects on the baryon-strangeness correlation in quark combination models  }

\author{Feng-lan Shao}
\affiliation{School of Physics and Engineering, Qufu Normal University, Shandong 273165, China}

\author{Jun Song}
\affiliation{Department of Physics, Jining University, Shandong 273155, China}

\author{Rui-qin Wang}
\affiliation{School of Physics and Engineering, Qufu Normal University, Shandong 273165, China}

\begin{abstract}
The baryon-strangeness correlation in the hadronization of the quark matter is studied within the quark combination mechanism. We calculate the correlation coefficient $C_{BS} = -3\big(\langle B S \rangle -\langle B\rangle \langle S\rangle\big)/\big( \langle S^2 \rangle-\langle S \rangle^2 \big)$ of initial hadrons produced from the deconfined free quark system with $C^{(q)}_{BS}=1$. The competition of the production of baryons against that of mesons is the key dynamics that is most relevant to the change of baryon-strangeness correlation during system hadronization.  Results of quark combination under the Poisson statistics agree with the statistical model predictions for a hadron resonance gas at vanishing chemical potential but differ from those at relatively large chemical potentials. Results beyond Poisson statistics are also obtained and are compared with calculations of Lattice QCD in the phase boundary, giving the best agreement at temperature $T =163$ MeV. We predict the dependence of the $C_{BS}$ of hadron system on the baryon chemical potential and strangeness.  These predictions are expected to be tested by the future Lattice QCD calculations at non-zero chemical potentials and/or by the Beam Energy Scan experiment of STAR Collaboration at RHIC.
\end{abstract}
\maketitle
\section{introduction}

Baryon-strangeness correlation is an effective diagnostic tool for the relevant degrees of freedom of the hot nuclear matter produced in relativistic heavy ion collisions \cite{koch05cbs}. It has been extensively investigated by various phenomenological models of high energy collisions \cite{cbsByURQMD,BMuller06bs,qMD08cbs,mayg08cbs,Mayg10cbs,PQM12cbs} and first-principle calculations in Lattice QCD \cite{lqcd06cbs,lqcdCBS}. The correlation is usually quantified by \cite{koch05cbs}
\begin{equation}
  C_{BS} \equiv -3 \frac{\langle B S \rangle -\langle B\rangle \langle S\rangle} {\langle S^2 \rangle -\langle S \rangle^2} =-3 \frac{\langle B S \rangle} {\langle S^2 \rangle}.
  \label{def_cbs}
\end{equation}
Here, angle brackets denote the event/ensemble average. In the second step, strangeness neutrality $\langle S\rangle =0$ is applied to relativistic heavy ion collisions.
For a deconfined system consisting of free quarks and antiquarks, $C^{(q)}_{BS} =1$ because strangeness is carried only by the strange (anti-)quarks which carry the baryon number in strict proportion to their strangeness with the coefficient -1/3, i.e., $B_s = -\frac{1}{3}S_s$.
In contrast, the relation between baryon number and strangeness in a hadron system is multiple, e.g., baryon number 1 for a strange baryon but 0 for a strange meson. $C_{BS}$ of hadron system is usually smaller than one (at low baryon number density) due to the fact that most strange quarks, at hadronization, will come into mesons instead of baryons which unlock the intimate baryon-strangeness correlation existed previously in quarks.
Statistical model estimation of $C_{BS}$ for a hadron resonance gas at zero baryon number density is about 0.66 \cite{koch05cbs}. Calculations of Lattice QCD also show that the $C_{BS}$ of the strong-interacting system at temperature above the phase transition temperature $T_c$ tends to one while near and below $T_c$ the system $C_{BS}$ decreases rapidly \cite{lqcd06cbs,lqcdCBS}.

Hadronization refers to the process of the formation of hadrons out of quarks and/or gluons, accompanying with the change of the correlation between baryon number and strangeness. Phenomenological models describing the hadronization should reproduce this change of baryon-strangeness correlation due to the transformation of the degrees of freedom in the system.
Quark combination is one of the effective mechanisms for the hadronization of the hot quark matter produced in relativistic heavy ion  collisions and has explained lots of experimental phenomena in heavy ion collisions at SPS, RHIC and recent LHC, see, e.g. Refs\cite{Fries22003prl,Greco2003prl,hwa04,sdqcm,Shuryak10qcm,wqr12}. The change of baryon-strangeness correlation during quark combination hadronization is intuitive. A strange quark(antiquark) combines with a light antiquark(quark) to form a strange meson, which completely destroys the original baryon-strange correlation carried by strange (anti)quark, i.e., $B_{(s\bar{q})}=0\times S_{(s\bar{q})}$,  but inherits the strangeness lossless. Occasionally, a strange quark combines with a strange antiquark to form a hidden-strange meson, which inherits nothing from original quarks. On the contrary, in the baryon formation a strange quark combines with two light quarks to form a baryon, which alters the baryon-strangeness correlation coefficient $B_{(sqq)}=-S_{(sqq)}$. The combination of two strange quarks with one light quark also alters the correlation coefficient $B_{(ssq)}=-\frac{1}{2}S_{(ssq)}$. Obviously, the $C_{BS}$ of the system depends on the relative proportion of the produced baryons to mesons.

In this paper, we study the change of system $C_{BS}$ caused by hadronization in quark combination mechanism (QCM).  We discuss in details how the dynamics of baryon-meson production competition at hadronization to dominate the $C_{BS}$ of hadron system.
In addition, we study the dependence of $C_{BS}$ on the strangeness content and the baryon number density of the system, which is related to the hot quark system produced in relativistic heavy ion collisions at different collisional energies. We compare our results with the calculations of Lattice QCD \cite{lqcd06cbs,lqcdCBS} and the prediction of statistical model for a hadron resonance gas \cite{koch05cbs}.

The paper is organized as follows. In Sec.~II, we present a working model in QCM for the yields of hadrons. In Sec.~III, using the model we explain the experimental data of the mid-rapidity yields of strange hadrons that are most relevant to $C_{BS}$ calculations in central Pb+Pb collisions at $\sqrt{s_{NN}}=2.76$ TeV. In Sec.~IV, we calculate the $C_{BS}$ of the hadron system in which two assumptions are made for initially produced hadrons. One assumption is that different kinds of hadrons are uncorrelated, which can be expected in the case that the quark system existed previously is made up of free quarks and antiquarks ($C^{(q)}_{BS}=1$). In this paper, we only consider this kind of quark system since the effects of hadronization on system $C_{BS}$ are already addressed properly.  The other assumption is Poisson statistics for yield fluctuations of hadrons.  In Sec.~V, we present a different approach of calculating $C_{BS}$ which is beyond Poisson statistics and can also reflect intuitively the essence of baryon-strangeness correlation. 
 Summary and discussion are finally given in Sec.~VI.

\section{ A working model of hadron yields}

In this section, we present a working model of the yields of hadrons that are produced from the deconfined quark phase after hadronization in the framework of quark combination mechanism, for the facility of the subsequent correlation studies. As discussed in introduction, baryon-strangeness correlation $C^{(h)}_{BS}$ is strongly dependent on the relative production of baryons to mesons, which requires our model to well address this point. Therefore, we introduce the working model according to the following strategy: (1) give the global properties for the global production of all mesons, all baryons and all antibaryons, in which the production competition of baryons against  mesons is properly addressed. (2) on the basis of (1), give the yield formulas of various identified hadrons.

In previous work \cite{sjbbar13}, we have studied the properties of the global production of all mesons, baryons and antibaryons, and obtained their yield formulas utilizing only the basic ideas of QCM. We start from a deconfined system consisting of $N_q$  constituent quarks and $N_{\bar{q}}$ antiquarks. Here, gluons contribution to the system on the threshold of hadronization is replaced by pairs of quark and antiquark. After hadronization, the system changes the basic degrees of the freedom to become the hadronic system and produces \emph{in average} $B(N_q,N_{\bar{q}})$ baryons, $\bar{B}(N_q,N_{\bar{q}})$ antibaryons and $M(N_q,N_{\bar{q}})$ mesons.
Four properties of hadron production from general principles are used to constrain the behavior of their yield formulas.

\noindent (1) charge conjugation symmetry of the hadron yields,
\begin{equation}
\begin{split}
B(N_q,N_{\bar{q}}) =\bar{B}(N_{\bar{q}},N_q),  \\
M(N_q,N_{\bar{q}}) = M(N_{\bar{q}},N_q).
\end{split}
\label{exchange}
\end{equation}
(2)  unitary of the hadronization, i.e. production of mesons, baryons and antibaryons should exhaust all quarks and antiquarks of the system existed previously
\begin{equation}
\begin{split}
   M(N_q,N_{\bar{q}})+3B(N_q,N_{\bar{q}})=N_q ,\\
   M(N_q,N_{\bar{q}})+3\bar{B}(N_q,N_{\bar{q}})=N_{\bar{q}}.
\end{split}
\label{unitarity}
\end{equation} (3) boundary condition, i.e.,
\begin{equation}
\begin{cases}
  \bar{B}=0, \ \  B=\frac{N_q}{3}, M=0         & \text{if } N_{\bar{q}}=0 \\
  \bar{B}=\frac{N_{\bar{q}}}{3},   B=0, \  \ M=0 &\text{if } N_{q}=0
\end{cases}
\label{boundary}
\end{equation}
(4) linear response of meson and baryon yields to quark antiquark numbers
\begin{equation}
\begin{split}
   M(\lambda N_q,\lambda N_{\bar{q}})=\lambda   M(N_q,N_{\bar{q}}), \\
   B(\lambda N_q,\lambda N_{\bar{q}})=\lambda  B(N_q,N_{\bar{q}}), \\
   \bar{B}(\lambda N_q,\lambda N_{\bar{q}})=\lambda  \bar{B}(N_q,N_{\bar{q}}).
\end{split}
\label{linearity}
\end{equation}
Using these properties, we obtained in Ref.\cite{sjbbar13} the yield formulas of baryons, antibaryons and mesons,
\begin{eqnarray}
  M(x,z) &=&  \frac{x}{2} \{  1- z \frac{(1+z)^{a} + (1-z)^{a} }{(1+z)^{a}-(1-z)^{a}}  \}, \nonumber \\
  B(x,z) &=& \frac{x\,z}{3}\frac{(1+z)^{a}}{(1+z)^{a}-(1-z)^{a}}, \label{bmpara} \\
  \bar{B}(x,z)&=& \frac{x\,z}{3}\frac{(1-z)^{a}}{(1+z)^{a}-(1-z)^{a}}. \nonumber
\end{eqnarray}
Here, we have rewritten $x= N_q + N_{\bar{q}}$ which characterizes the bulk property of the system related to the system size or energy and rewritten $z=(N_q -N_{\bar{q}})/x$ which depicts the asymmetry between quarks and antiquarks in the system ($|z|\le1$) and is a measurement of the baryon number density of the system.

The production competition of baryons against mesons is often quantified by the yield ratios $R_{B/M}(z)=B(x,z)/M(x,z)$ and $R_{\bar{B}/M}(z)=\bar{B}(x,z)/M(x,z)$, which are the function of only $z$.
The factor $a$ in Eq.~(\ref{bmpara}) represents the degree of the baryon-meson competition by the relation $a=\frac{1}{3 R_{B/M}(0)}+1$.
Studies in Ref. \cite{wqr12} show the $R_{B/M}(0)$ of value about $1/12$, i.e., $a \approx 5$, can well describe yield ratios of various baryons to mesons in heavy ion collisions at LHC energy. We note that such a competition as well as yield formulas Eq.~(\ref{bmpara}) can be properly addressed by a phenomenological combination rule in the quark combination model developed by Shandong group (SDQCM) \cite{sdqcm}.
In addition, using vested $R_{B/M}(0)$, we have successfully explained the yield ratios of various anti-hadrons to hadrons at non-zero $z$ region in relativistic heavy ion collisions \cite{sjbbar13,wqr12,wqr14}, i.e., the data of these yield ratios at different collision energies and at different rapidities.
We emphasis that through the dependence of $C^{(h)}_{BS}$ on the factor $a$ or $R_{B/M}(0)$ we can address the effects of hadronization on the baryon-strangeness correlation of the system.

To study $C^{(h)}_{BS}$, we have to obtain the yield formulas of identified hadron. Given the total yield of baryon, that of antibaryon and that of meson, the inclusive/averaged yields of identified hadrons $M_{i}(q_1\bar{q}_2)$, $B_{j}(q_1q_2q_3)$ are calculated by their individual production weights,
\begin{eqnarray}
  N_{M_i} &=& P_{M_i} M(x,z) = C_{M_i} P_{q_1\bar{q}_2,M} M(x,z)   \label{miformula}, \\
  N_{B_j} &=& P_{B_j} B(x,z) = C_{B_j} P_{q_1q_2q_3,B} B(x,z).     \label{bjformula}
\end{eqnarray}
Here, $P_{q_1\bar{q}_2,M}$ denotes the probability that, as a meson is known to be produced, the flavor content of this meson is $q_1\bar{q}_2$.  $C_{M_i}$ further denotes the branch ratio of this meson with given flavor composition $q_1\bar{q}_2$ to a specific meson state $M_i$.  $C_{M_i} P_{q_1\bar{q}_2,M}$ thus gives the emerging probability of a specific meson, $P_{M_i}$, when a meson is known to be formed.
Similarly, $P_{q_1q_2q_3,B}$ denotes the probability that, as a baryon is known to produced, the flavor content of this baryon is $q_1q_2q_3$, and $C_{B_j}$ denotes the branch ratio of this baryon with given flavor composition $q_1q_2q_3$ to a specific baryon state $B_j$, and $C_{B_j} P_{q_1q_2q_3,B}$ thus gives the emerging probability of a specific baryon, $P_{B_j}$, when a baryon is known to be formed. A similar formula holds for antibaryons.

The probability $P_{q_1\bar{q}_2,M}$ is posteriorly evaluated by the proportion of $q_1 \bar{q}_2$ pairs in all quark-antiquark pairs in the system, $P_{q_1\bar{q}_2,M} = N_{q_1\bar{q}_2}/N_{q\bar{q}}$, by considering the fact that every quark/antiquark has both the probability of entering into a meson and the probability of entering a baryon/antibaryon. Here, $N_{q}=\sum_{i} N_{q_i}$ and $N_{\bar{q}}=\sum_{i} N_{\bar{q}_i}$ are total number of quarks and antiquarks in the system, respectively. We have $N_{q_1 \bar{q_2}}= N_{q_1 }N_{\bar{q_2}}$ and $N_{q\bar{q}} =  N_{q }N_{\bar{q}}$.
Similarly, the probability $P_{q_1q_2q_3,B}$ is evaluated by the proportion of $q_1q_2q_3$ combinations to all three-quark combinations in the system, $P_{q_1q_2q_3,B}=N_{iter} N_{q_1q_2q_3}/N_{qqq}$. $N_{q_1 q_2 q_3}$ is the number of $q_1 q_2 q_3$ combination which satisfies $ N_{q_1 q_2 q_3}=N_{q_1}N_{q_2}N_{q_3}$ for $q_1\neq q_2 \neq q_3$, $ N_{q_1 q_2 q_3}=N_{q_1}(N_{q_1}-1)N_{q_3}$ for $q_1 = q_2 \neq q_3$ and $ N_{q_1 q_2 q_3}=N_{q_1}(N_{q_1}-1)(N_{q_1}-2)$ for $q_1= q_2 = q_3$.  $N_{qqq} =N_{q}(N_{q}-1)(N_{q}-2)$ is the number of all $qqq$ combination in the system.  In large quark number limit we can apply $N_{qqq} \approx N_{q}^3$ and $N_{q_1 q_2 q_3}\approx N_{q_1}N_{q_2}N_{q_3}$ always.
$N_{iter}$ stands for the number of possible iterations of $q_1 q_2 q_3$, which is 1, 3, and 6 for three identical flavor, two different flavor, and three different flavor cases, respectively.

$C_{M_j}$ and/or $ C_{B_j}$ denote the branch ratio of a given flavor composition (e.g., $u\bar{s}$) to a specific hadron state (e.g., $K^{+}$) under the condition that they are known form a hadron. In the case that when the ground state $J^P=0^-$ and $1^-$ mesons and $J^P=\frac{1}{2}^+$ and $\frac{3}{2}^+$ baryons are considered only, we have, for mesons,
\begin{equation}
   C_{M_j} =  \left\{
   \begin{array}{ll}
        {1}/{(1+R_{V/P})}~~~~~~~~   \textrm{for } J^P=0^-  \textrm{ mesons},  \\
                 {R_{V/P}}/{(1+R_{V/P})}~~~~         \textrm{for } J^P=1^-  \textrm{ mesons},
   \end{array} \right.  \nonumber
\end{equation}
where $R_{V/P}$ represents the ratio of the $J^P=1^-$ vector mesons to the $J^P=0^-$ pseudoscalar mesons of the same flavor composition; and for baryons,
\begin{equation}
C_{B_j} =  \left\{
    \begin{array}{ll}
        {R_{O/D}}/{(1+R_{O/D})}~~~~   \textrm{for } J^P=({1}/{2})^+  \textrm{ baryons},  \\
        {1}/{(1+R_{O/D})}~~~~~~~~         \textrm{for } J^P=({3}/{2})^+  \textrm{ baryons},
    \end{array} \right. \nonumber
\end{equation}
except that $C_{\Lambda}=C_{\Sigma^0}={2R_{O/D}}/{(1+2R_{O/D})}$, $C_{\Sigma^{*0}}={1}/{(1+2R_{O/D})}$, $C_{\Delta^{++}}=C_{\Delta^{-}}=C_{\Omega^{-}}=1$ .
Here, $R_{O/D}$ stands for the ratio of the $J^P=(1/2)^+$ octet to the $J^P=(3/2)^+$ decuplet baryons of the same flavor composition.  The two parameters $R_{V/P}$ and $R_{O/D}$ can be determined using the data from different high energy reactions \cite{sdqcm,wq95}, and they are taken to be 3 and $2$ in this paper, respectively.

\section{yields of strange hadrons at LHC}

With above working model, we can conveniently predict yields of various identified hadrons.  As an illustration, we now explain the experimental data of mid-rapidity yields of strange hadrons $K$, $\Lambda$, $\Xi$ and $\Omega^{-}$ in Pb+Pb collisions at $\sqrt{s_{NN}}=2.76$ TeV. These hadrons are most relevant to $C_{BS}$ calculation.

As applying yield formulas in Sec.~II to the finite rapidity window in heavy ion collisions, we note that Eqs.~(\ref{exchange}) and (\ref{boundary}-\ref{linearity}) still hold generally and Eq.~(\ref{unitarity}) also approximately holds with good numerical accuracy due to the locality of the hadronization, and therefore no conceptual issues exist for our mid-rapidity predictions. In addition, our formulas in Sec.~II give the averaged yields of hadrons at the given quark system with the fixed quark numbers while the experimental data of hadronic yields are event-averaged quantities. The produced quark system in heavy ion collisions at a given collision energy is varied in size event-by-event and thus the number of quarks and that of antiquarks should follow a certain distribution around the averaged quark number $\langle N_{q_i}\rangle$ and antiquark numbers $\langle N_{\bar{q}_i} \rangle$ (where $i=u, d, s$ considered in this paper). Our final predictions of hadron yields should be the average over this distribution. In general such averages depend on the precise form of the distribution. Here, we approximate these averages by taking the corresponding values of the quantities at the event averages $\langle N_{q_i}\rangle$ and $\langle N_{\bar{q}_i} \rangle$, i.e.,
\begin{eqnarray}
  \langle N_{M_i}\rangle &=&  C_{M_i} \overline{P}_{q_1\bar{q}_2,M} M(\langle x\rangle,\langle z \rangle)   \label{miave}, \\
  \langle N_{B_j}\rangle &=& C_{B_j} \overline{P}_{q_1q_2q_3,B} B(\langle x\rangle,\langle z \rangle),     \label{bjave}
\end{eqnarray}
where $\langle x \rangle = \langle N_q\rangle  + \langle N_{\bar{q}} \rangle$, $\langle x \rangle \langle z \rangle = \langle N_q \rangle - \langle N_{\bar{q}} \rangle$ and $\overline{P}$ is also calculated with the event averaged quark numbers. Such approximation can be expected in considering that properties Eqs.~(\ref{exchange} - \ref{linearity}) in our derivation of hadron yields are also apparently satisfied even for the event-averaged quantities.

For convenience, we use factors $\lambda_s = \langle N_{\bar{s}}\rangle / \langle N_{\bar{u}} \rangle = \langle N_{\bar{s}}\rangle / \langle N_{\bar{d}}\rangle $ and associated $\lambda'_s = \langle N_{s}\rangle /\langle N_{u}\rangle =\langle  N_{s}\rangle /\langle N_{d}\rangle $ to denote the suppression of strange anti-quarks relative to light antiquarks and that of strange quarks to light quarks, respectively,  and simply the yield formulas of hadrons. Here, isospin symmetry between (anti-)up quark and (anti-)down quark is applied.

Considering the decay contribution from the short-lived resonances, we obtain the yields of final state hadrons
\begin{equation}
\langle N_{h_j}^{(f)}\rangle  = \langle N_{h_j}\rangle + \sum_{k} Br( h_k \to h_j) \ \langle N_{h_k}\rangle ,
\end{equation}
where we use the superscript $(f)$ to denote the results for the final hadrons to differentiate them from those for the directly produced hadrons.
The data of the decay branch ratios are taken from the PDG \cite{PDG}.

With $\lambda_s = \lambda'_s$ at LHC ($\langle z\rangle =0$ approximation), we obtain yields of these strange hadrons in final state
\begin{eqnarray}
 \langle N^{(f)}_{K^+}\rangle  &=&  \langle N_{K^{+}}\rangle  + \langle N_{K^{*+}}\rangle  Br(K^{*+} \to K^+) + \langle N_{K^{*0}}\rangle  Br(K^{*0} \to K^+) \nonumber \\
&=& \Big( 1+ 0.493\frac{R_{V/P}}{1+R_{V/P}}\lambda_s \Big) \frac{\lambda_s}{(2+\lambda_s)^2} M(\langle x \rangle,0), \label{kyield} \\
\langle N^{(f)}_{\Lambda}\rangle  &=& \langle N_{\Lambda}\rangle  + \langle N_{\Sigma^0}\rangle Br(\Sigma^0 \to \Lambda) + \langle N_{\Sigma^{*+}}\rangle Br(\Sigma^{*+}\to \Lambda) \nonumber \\
	&{}& + \  \langle N_{\Sigma^{*0}}\rangle Br(\Sigma^{*0}\to \Lambda)+ \langle N_{\Sigma^{*-}}\rangle Br(\Sigma^{*-}\to \Lambda)  \nonumber \\
&=& \frac{7.736 \lambda_s}{(2+\lambda_s)^3} B(\langle x\rangle,0), \\
N^{(f)}_{\Xi^{-}} &=& \langle N_{\Xi^{-}}\rangle + \langle N_{\Xi^{*-}}\rangle Br(\Xi^{*-}\to \Xi^{-})+ \langle N_{\Xi^{*0}}\rangle Br(\Xi^{*0}\to \Xi^{-})\nonumber \\
&=& 3 \frac{\lambda_s^2}{(2+\lambda_s)^3} B(\langle x\rangle,0), \\
\langle N^{(f)}_{\Omega^-}\rangle &=& \langle N_{\Omega^{-}}\rangle = \frac{\lambda_s^3}{(2+\lambda_s)^3} B(\langle x\rangle,0).\label{Omegayied}
\end{eqnarray}
Here, we have only taken into account the strong and electric-magnetic (S\&EM) decays of short-lived resonances.

\begin{figure}[!htp]
\centering
\includegraphics[width=\linewidth]{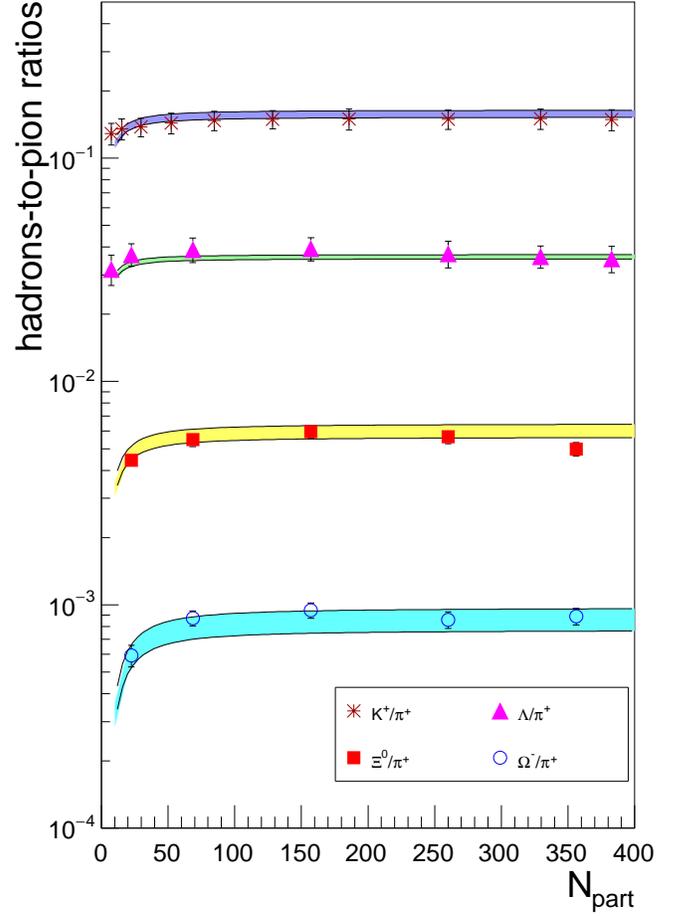}
\caption{(Color online)  Yield ratios $K^{+}/\pi^{+}$, $\Lambda/\pi^+$, $\Xi^{-}/\pi^+$ and $\Omega^{-}/\pi^+$ at mid-rapidity as the function of nuclear participants $N_{part}$ in central Pb+Pb collisions at $\sqrt{s_{NN}}=2.76$ TeV. Symbols are experimental data from Refs.\cite{aliceKpion13,aliceKLam13,aliceXiOmg13}. Shadow regions are our results with a $N_{part}$-dependent strangeness $\lambda_s(N_{part}) = (0.43\pm0.02) /(1+10.5 N_{part}^{-1.3})$. }
\label{fig1}
\end{figure}

Absolute yields of these hadrons are dependent on the total quark number $\langle x\rangle$ of the system by $M(\langle x \rangle,0)=2\langle x \rangle /5$ and $B(\langle x\rangle ,0)=\langle x\rangle /30$ with fixed $R_{B/M}(0)=1/12$. To eliminate this $\langle x \rangle$ dependence, we consider the relative production of these hadrons to pions, i.e., the yield ratios of these hadrons to pions, which finally rely only on the strangeness of the system, beside of the baryon-meson competition in hadronization. Since the decay contributions to pion are complex, we here give directly the numerical result of the yield of $\pi^+$, i.e., $\langle N^{(f)}_{\pi^+}\rangle = 0.213 \langle x \rangle$ under S\&EM decays, instead of the detailed compositions like Eqs.~(\ref{kyield}-\ref{Omegayied}).
In Fig.~\ref{fig1}, we use Eqs.~(\ref{kyield}-\ref{Omegayied}) to explain the experimental data of the mid-rapidity yield ratios $K^{+}/\pi^{+}$, $\Lambda/\pi^+$, $\Xi^{-}/\pi^+$ and $\Omega^{-}/\pi^+$ in central Pb+Pb collisions at $\sqrt{s_{NN}}= 2.76$ TeV \cite{aliceKpion13,aliceKLam13,aliceXiOmg13}. To incorporate the change of $\lambda_s$ at different collision centralities, a varied strangeness $\lambda_s(N_{part}) = (0.43\pm0.02) /(1+10.5 N_{part}^{-1.3})$ is used.
We can see that the hierarchy properties in yields of these strange hadrons and their $N_{part}$ dependence can be systematically described by our formulas Eqs.~(\ref{kyield}-\ref{Omegayied}).
Here, we would like to emphasis that the yield difference between $K$ and those hyperons is mainly due to the baryon-meson competition at hadronization while the hierarchy structures among $\Lambda$, $\Xi^-$ and $\Omega^{-}$ are mainly strangeness relevant.

Applying our yield formulas to other hadrons at LHC and those at RHIC energies with nonzero baryon number densities, we also find a good agreement with available experimental data. This is not surprise.
In fact, QCM has already shown its effectiveness in explaining the data of hadronic yields and longitudinal rapidity distributions in relativistic heavy ion collisions at different collisional energies \cite{biro1999,alcor2000,alcorC2000,sdqcm,sj09ijmpa,wqr12,lsSLX12}. The related low-$p_T$ issues of QCM such as entropy conservation and pion production have been properly addressed \cite{biro1999,ckm03,nonaka05,hwa04,biro2007entropy,fries08review,SJentrop10}.  There are also many successful applications of QCM on correlation studies, e.g.,~multi-hadron yield correlations \cite{wqr12,sjbbar13,wqr14,wqr15charm}, baryon-meson correlated emission \cite{fries:2004hd,fries:2007ix} as well as the charge balance function \cite{Bialas04,SJ12FB}.

\section{$C_{BS}$ of hadrons under Poisson fluctuations}

For initial hadrons produced by the hadronization of deconfined quark system, the baryon number of the system is $B=\sum_\alpha Q_{\alpha,B} N_{\alpha}$ and the strangeness $S = \sum_\alpha Q_{\alpha,S} N_{\alpha}$, where the species $\alpha$ has baryon number $Q_{\alpha,B}$ and strangeness $Q_{\alpha,S}$.
By definition Eq.~(\ref{def_cbs}), the $C_{BS}$ of hadrons is
\begin{equation}
	C_{BS}^{(h)} = -3 \frac{ \sum_{\alpha,\beta} Q_{\alpha,B} Q_{\beta,S} C_{\alpha\beta} }{ \sum_{\alpha,\beta} Q_{\alpha,S} Q_{\beta,S} C_{\alpha\beta}},
\end{equation}
where the covariance $C_{\alpha \beta} = \langle N_{\alpha} N_{\beta} \rangle -\langle N_{\alpha} \rangle \langle N_{\beta} \rangle$ describes the correlation between hadron $\alpha$ and hadron $\beta$. A general calculation of $C_{\alpha \beta}$ in QCM is still unavailable in the current progress of hadronization phenomenology since there still has lots of unsolved dynamics in hadronization due to its non-perturbative feature. On all ``on market'' combination models, there are a few ones that can give the calculation of $C_{\alpha \beta}$ with their own specific model details/assumptions.  Such specific calculations are not the purpose of this paper since we intend to analyze the hadronization effects in a general and transparent way. Here, we consider a simple case that after hadronization different kinds of produced hadrons are uncorrelated, i.e. $C_{\alpha \beta}=\delta_{\alpha \beta} \sigma^{2}_{\alpha}$. This can be expected if the quark system existed previously is made up of free quarks and antiquarks (i.e.,~$C^{(q)}_{BS}=1$). We note that, above $T_c$, the $C_{BS}$ of strong-interacting system calculated by Lattice QCD \cite{lqcdCBS} indeed tends to one and the off-diagonal flavor susceptibilities also tend to be relatively small. We only consider this kind of quark system in this paper because the effects of hadronization on system $C_{BS}$ are already addressed clearly .

In this section, we first calculate the $C_{BS}$ of the initial hadron system, assuming the Poisson statistics $\sigma^{2}_{\alpha} \approx \langle N_{\alpha} \rangle$  for the yield distribution of identified hadrons. Now we have
\begin{equation}
	C^{(h)}_{BS} = -3 \frac{ \langle B S \rangle }{ \langle S^2 \rangle } \approx -3 \frac{ \sum_\alpha Q_{\alpha,B} Q_{\alpha,S} \langle N_\alpha \rangle }{ \sum_\alpha Q^2_{\alpha,S} \langle N_\alpha \rangle }. \label{PoissonCbs}
\end{equation}
With Eqs.~(\ref{miave}) and (\ref{bjave}), we are ready to calculate $C_{BS}$ of hadrons via Eq.~(\ref{PoissonCbs}),
\begin{eqnarray}
 \langle B S\rangle  &= &- \{ \langle N_{\Lambda}\rangle +\langle N_{\Sigma^{\pm,0}}\rangle + \langle N_{\Sigma^{*\pm,0}}\rangle \}  -2 \{\langle N_{\Xi^{0,-}}\rangle + \langle N_{\Xi^{* 0,-}}\rangle \} \nonumber \\
	                  &{} & - 3 \langle N_{\Omega^{-}}\rangle - anti-hyperons \nonumber \\
 = &{-}& \frac{12\lambda'_s + 12 \lambda'^{2}_s + 3 \lambda'^{3}_s}{(2+\lambda'_s )^3} B(\langle x \rangle ,\langle z \rangle)- \frac{12\lambda_s + 12 \lambda^{2}_s + 3 \lambda^{3}_s}{(2+\lambda_s )^3} \nonumber \\
 &{\times}& \bar{B}(\langle x \rangle ,\langle z \rangle),
\end{eqnarray}
and
\begin{eqnarray}
  \langle S^{2} \rangle &= & \{\langle N_{\Lambda}\rangle  + \langle N_{\Sigma^{\pm,0}}\rangle + \langle N_{\Sigma^{*\pm,0}}\rangle \}  + 4 \{ \langle N_{\Xi^{0,-}}\rangle + \langle N_{ \Xi^{* 0,-}}\rangle  \} \nonumber \\
	  &{}  & + 9 \langle N_{\Omega^{-}}\rangle  + anti-hyperons  \nonumber \\
&{} &+ \{\langle K^{\pm}\rangle  + \langle K^{*\pm}\rangle  + \langle K^{0}\rangle  + \langle K^{*0}\rangle + \langle \bar{K^{0}}\rangle + \langle \bar{K}^{*0}\rangle \} \nonumber \\
   &= &\frac{12\lambda'_s + 24 \lambda'^{2}_s + 9 \lambda'^{3}_s}{(2+\lambda'_s )^3} B(\langle x \rangle ,\langle z \rangle ) + \frac{12\lambda_s + 24 \lambda^{2}_s + 9 \lambda^{3}_s}{(2+\lambda_s )^3} \nonumber \\
  &{}& \times \bar{B}(\langle x \rangle,\langle z \rangle) 	+ \frac{2\lambda_s + 2\lambda'_s}{(2+\lambda_s)(2+\lambda'_s)}M(\langle x \rangle ,\langle z \rangle).
\end{eqnarray}
We note that above correlations are independent of the parameters $R_{O/D}$ and $R_{V/P}$ and thus they are unaffected by S\&EM decays.
Substituting them into Eq.~(\ref{PoissonCbs}), we obtain
\begin{equation}
  C^{(h)}_{BS} = 3 \frac{ \frac{3\lambda'_s}{2+\lambda'_s}R_{B/M}(\langle z \rangle) + \frac{3\lambda_s}{2+\lambda_s}R_{\bar{B}/M}(\langle z \rangle)  }{ \frac{3\lambda'_s(3\lambda'_s+2) }{(2+\lambda'_s)^2 }R_{B/M}(\langle z \rangle)  +\frac{3\lambda_s(3\lambda_s+2) }{(2+\lambda_s)^2 }R_{\bar{B}/M}(\langle z \rangle)  + \frac{2\lambda'_s +2\lambda_s }{ (2+\lambda'_s)(2+\lambda_s)}  },
  \label{cbs_0_full}
\end{equation}
which gives the dependence of $C^{(h)}_{BS}$ on baryon-meson competition factor $R_{B/M}(0)$, strangeness $\lambda_s$ and the baryon number density of the system.

We firstly consider the situation of zero baryon number density $\langle z \rangle=0$ to study the dependence of $C^{(h)}_{BS}$ on  $R_{B/M}(0)$ and $\lambda_s$. With $\lambda'_s = \lambda_s$ and $R_{B/M}(0)=R_{\bar{B}/M}(0)$, we get
\begin{equation}
  C^{(h)}_{BS} = 3 \frac{ (2+\lambda_s)R_{B/M}(0) } {  (3\lambda_s +2)R_{B/M}(0) + 2/3 }.
\end{equation}
Fig.~2 (a) shows the dependence of $C^{(h)}_{BS}$ on the strangeness $\lambda_s$, as the $R_{B/M}(0)$ is taken to 1/12. One can see that $C^{(h)}_{BS}$ is insensitive to the change of the strangeness, i.e. $C^{(h)}_{BS}$ only increases about 5\% as $\lambda_s$ increases from 0.3 (the rough value in $pp$ reactions) to 0.7  (almost maximum value occurred in heavy ion collisions). In addition, we see that $C^{(h)}_{BS}$ of initial hadron system is obviously small than one (the value of ideal quark system) because the produced strange mesons significantly outnumber the strange baryons in the current baryon-meson competition $R_{B/M}(0)=1/12$.
Fig.~2 (b) shows the dependence of $C^{(h)}_{BS}$ on baryon-meson competition factor $R_{B/M}(0)$, as the strangeness $\lambda_s$ is taken to the saturated value 0.43 in relativistic heavy ion collisions. Here, the saturated $\lambda_s$ is extracted from the fit of yields of strange hadrons in Sec. III and in our previous work \cite{wqr12}, and we note that this value is consistent with calculations of Lattice QCD at $T_c$ \cite{lqcd06cbs,lsLQCD06}. Clearly, we see that the increase of $R_{B/M}(0)$ will enhance the yields of baryons against mesons and the $C^{(h)}_{BS}$ increases rapidly. With preferred values $R_{B/M}(0)=1/12$ and $\lambda_s =0.43$, $C^{(h)}_{BS}$ is about 0.65, much smaller than $C^{(q)}_{BS}=1$. To reach the unit correlation for hadrons, an extremely high baryon-meson competition factor $R_{B/M}(0)\approx 1/5$ is needed, which is completely unable to reproduce yields of strange mesons and baryons.

\begin{figure}[!htp]
\centering
  \includegraphics[width=\linewidth]{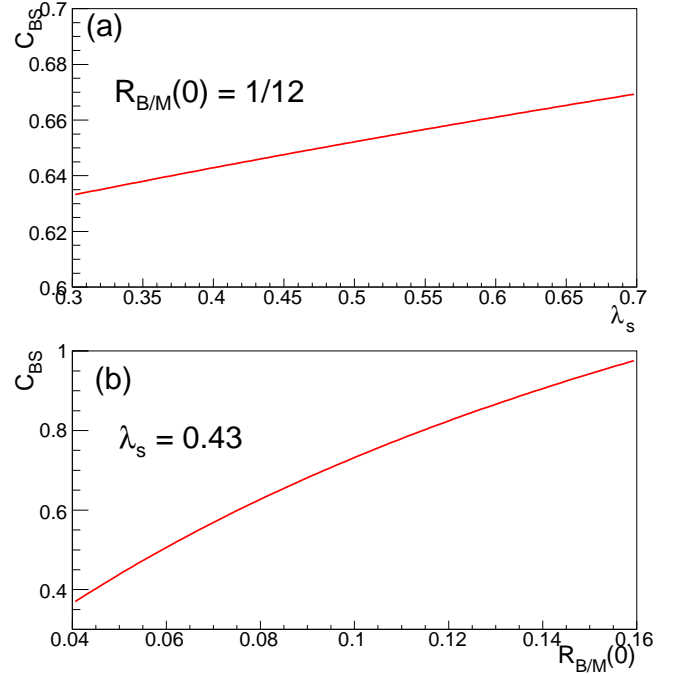}
  \caption{(Color online) (a) The dependence of $C^{(h)}_{BS}$ on strangeness suppression factor $\lambda_s$; (b) on baryon-meson competition factor $R_{B/M}(0)$. }
  \label{Cbsd1LsRbm}
\end{figure}

Subsequently, we study the dependence of $C^{(h)}_{BS}$ on the baryon number density of the system. In previous discussions, we use the quark-antiquark asymmetry $\langle z \rangle $ to characterize the baryon number density of the system. To compare our results with existed predictions of statistical models, we alternatively use the chemical potential $\mu_B$ which relates $\langle z \rangle$ via
\begin{equation}
\langle z \rangle = \frac{ 2 \sinh(\frac{\mu_B}{3T_c}) }{ 2\cosh(\frac{\mu_B}{3T_c}) + \lambda_s \exp(-\frac{\mu_B}{3T_c})},
\label{ztomub}
\end{equation}
under the assumption of Boltzmann distribution for thermalized quarks and antiquarks. Here, strangeness neutrality $N_s = N_{\bar{s}}$ is applied. $T_c$ is the temperature of the quark system at hadronization.

\begin{figure}[!htp]
\centering
  \includegraphics[width=\linewidth]{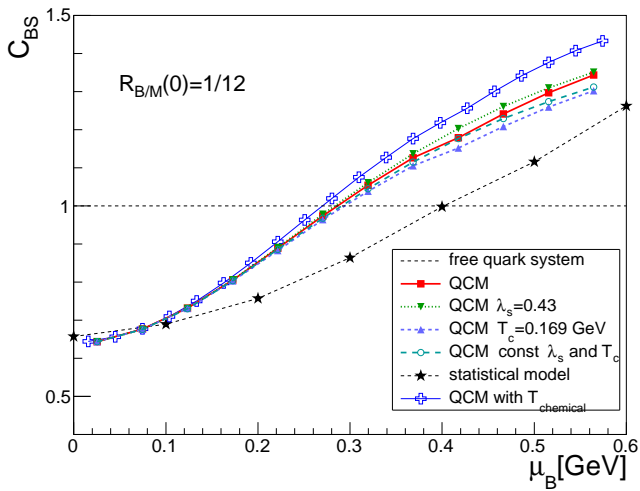}
  \caption{(Color online)  The dependence of $C_{BS}$ on the chemical potential $\mu_{B}$ of the system. The lines with filled squares, down-triangles, up-triangles and open circles are our results at both varied $\lambda_s$ and $T_c$, at constant $\lambda_s=0.43$, at constant $T_c = 0.169$ GeV, and at both constant $\lambda_s=0.43$ and $T_c=0.169 $ GeV, respectively. The solid line with open cross symbols is our results using the chemical freeze-out temperature. They are compared with the prediction of statistical model for hadron resonance gas \cite{koch05cbs}, the dashed line with stars.  }
  \label{cbsmub1}
\end{figure}

Fig.~\ref{cbsmub1} shows our predictions of the $C^{(h)}_{BS}$ of initial hadrons as the function of $\mu_B$. Here, we have taken into account the fact that both $T_c$ and $\lambda_s$ are varied with the $\mu_B$ for the hot quark matter produced in relativistic heavy ion collision. For the $\mu_B$ dependence of $T_c$, we apply the calculation of Lattice QCD by G.~Endr\H{o}di \emph{et al} \cite{tclqcd}, i.e., $T_c(\mu_B)=T_{c0} (1-0.0089 \mu_B^2/T_{c0}^2)$ with $T_{c0}$ being 0.169 GeV, the transition temperature at $\mu_B=0$. For the $\mu_B$ dependence of $\lambda_s$, we use our previous extractions at mid-rapidity in heavy ion collisions at different collision energies \cite{sdqcm,lsSLX12}, i.e., $\lambda_s=$ (0.43, 0.43, 0.44, 0.44, 0.48, 0.5, 0.57, 0.8, 0.7) as $\sqrt{s_{NN}}=$ (2760, 200, 130, 62.4, 17.3, 12.3, 8.7, 7.6, 6.3) GeV, and convert the energy dependence into $\mu_B$ dependence by the formula $\sqrt{s_{NN}}=(1.308 \text{GeV}/\mu_B -1)/0.273$ GeV in Refs.\cite{statMub11,statMub06}. The solid line with filled square symbols are our result.
We also calculate $C^{(h)}_{BS}$ at fixed $\lambda_s =0.43$ (dotted line with down-triangles), at fixed $T_c=0.169$ GeV (dashed line with up-triangles), and at both fixed $T_c =0.169$ GeV and fixed $\lambda_s=0.43$ (dashed line with open circles), respectively.

A striking behavior of $C^{(h)}_{BS}$ of initial hadrons is that it increases with the increasing $\mu_B$ and in the large $\mu_B$ region ($\mu_B \gtrsim 0.3$ GeV) it surpasses the unity and become higher at larger $\mu_B$. This is in sharp contrary to the free quark system existed previously where the correlation coefficient remains strictly unity at all temperatures and chemical potentials.
This is because that, as the $\mu_B$ increases, the relative production of baryons to mesons, i.e., $R_{B/M}(z)$ in our approach, becomes large, and then weight of this item in $C_{BS}$ formula Eq.~(\ref{cbs_0_full}) increases and thus $C_{BS}$ increases correspondingly.
Comparing the result of $C_{BS}$ at fixed $T_c$ with those with varied $T_c$, we see that the change(decrease) of $T_c$ at large chemical potential increases the $C_{BS}$ of the system about several percentages.
Comparing the result of $C_{BS}$ at fixed $\lambda_s$ with those with varied $\lambda_s$, we find that the change of the strangeness also slightly influences the $C_{BS}$ of the system.

We also compare our results with the early prediction of V.~Koch \emph{et al} \cite{koch05cbs} (the dotted line with star symbols) for a hadron resonance gas in the statistical treatment.
A $\mu_B$ dependent chemical freeze-out temperature, $T_{chemical}(\mu_B) \approx 0.17 - 0.13\mu^2_B -0.05\mu^4_B$ in Ref.\cite{shm03}, is used in their prediction, which falls obviously below the transition temperature we used above at intermediate and large $\mu_B$. For comparison, we also calculate the $C_{BS}$ of initial hadrons using their chemical freeze-out temperature and the result is presented in Fig.~\ref{cbsmub1} as the solid line with open cross symbols.
We find that our result is close to the prediction of statistical model at small $\mu_B$ but at intermediate and large $\mu_B$ region our result is about 15-20\% larger than that of statistical model.
We note that recently F.~Becattini \emph{et al} \cite{becattini13tc} have reconstructed  the original hadro-chemical equilibrium temperature after considering the effects of final hadron expansion phase and the results have closely followed the phase transition boundary line predicted by Lattice QCD \cite{tclqcd}.

\section{$C_{BS}$ of hadrons beyond Poisson statistics}

In above calculations of $C^{(h)}_{BS}$, Poisson statistics for hadronic yields is applied as an open approximation. Compared the obtained $C^{(h)}_{BS}$ at zero $\mu_B$ with calculations of Lattice QCD in the phase boundary \cite{lqcdCBS}, we find our result conform to Lattice calculations at $T = 150 \pm 3$ MeV. This value is lower than the usually expected deconfined temperature (around 165 MeV) \cite{lqcdCBS}, and is also lower than the chemical freeze-out temperature extracted from hadronic yields in statistical models which is about 166 MeV \cite{statMub06,statMub11}. Such an underestimation may be relevant to the applied Poisson statistics approximation since this application implies that a further sufficient evolution of hadron phase dominated by inelastic scatterings is probably needed. Therefore, in this section, we calculate the $C^{(h)}_{BS}$ of initial hadron system by another approach which is independent of the Poisson statistics assumption and also reflects intuitively the essence of the baryon-strangeness correlation.

To probe the intrinsic correlation between the baryon number and strangeness for a hadron system, we suppose that the system has a change by stochastically emitting a small amount of strange hadrons and then the system strangeness changes $\delta S^{(h)}$ and the baryon number also changes $\delta B^{(h)}$. Regarding the baryon number of the system $B^{(h)}$ as the function of the strangeness $S^{(h)}$, the change of baryon number due to the perturbation of strangeness can be calculated by
\begin{equation}
\delta B^{(h)} = \Big(\frac{\partial B^{(h)}_s }{ \partial S^{(h)}} \Big)\, \delta S^{(h)} + \mathcal{O}\big((\delta S^{(h)})^2\big),
\end{equation}
where $B^{(h)}_{s}$ is the baryon number carried by strange hadrons. To second order fluctuation of strangeness we get
\begin{equation}
C^{(h)}_{BS} = -3 \frac{\langle \delta B^{(h)} \delta S^{(h)}\rangle }{\langle {\delta S^{(h)}}^2 \rangle } \approx -3 \  \left. \frac{\partial B^{(h)}_s }{ \partial S^{(h)}} \right|_{\substack{\scriptscriptstyle \langle N_{h}\rangle }}.
\end{equation}
The partial derivative is evaluated at the event average values of hadron yields $\langle N_{h} \rangle$.
 Note that here we don't consider the correlations between light and strange hadrons induced by the interactions among hadrons, i.e., we consider an ideal hadron gas system after hadronization.

\begin{figure}[!htp]
\centering
  \includegraphics[width=\linewidth]{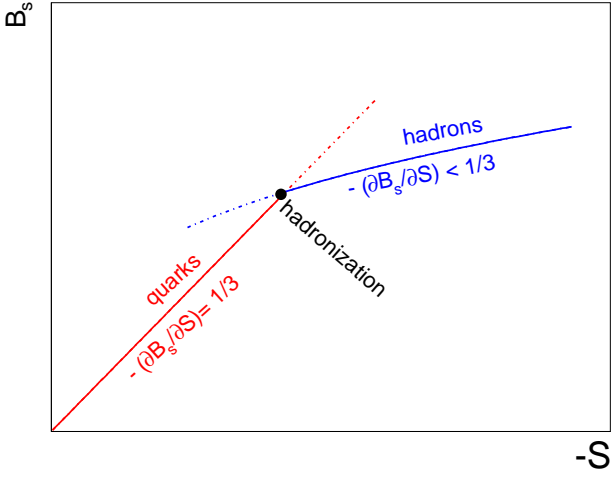}
  \caption{(Color online) Schematic picture for the transition of ideal quark phase to hadron phase by hadronization in $B-S$ plane. }
  \label{pbps}
\end{figure}

The same philosophy can be applied to the quark system before hadronization. Considering that the quark system is made up of free quarks and antiquarks with three flavors, the strangeness of the system $S^{(q)} = -(N_s - N_{\bar{s}})$ and the baryon number carried by these strange quarks and antiquarks  $B_s^{(q)} = \frac{1}{3}(N_s -N_{\bar{s}})$ and we get $C^{(q)}_{BS}=1$ exactly.
The difference of $C_{BS}$ between quarks and initial hadrons can be illustrated by the different slopes of two phases in $B$-$S$ plane, as shown in schematic Fig.~\ref{pbps}. The cross point between two phases stands for the hadronization that changes the basic degrees of freedom of the system and also stands for the global charge conservations during hadronization.

With yield formulas Eqs.~(\ref{miformula}) and (\ref{bjformula}), we have
\begin{eqnarray}
&&B^{(h)}_s = N_{\Lambda} + N_{\Sigma^{\pm,0}} +N_{\Sigma^{*\pm,0}} + N_{\Xi^{0,-}} +  N_{\Xi^{* 0,-}} + N_{\Omega^{-}}  \nonumber \\
&{}&\hspace{10pt}- \  anti-hyperons \\
&=& \frac{6 N_{uds} + 3 N_{uus} + 3 N_{dds} + 3 N_{uss} + 3 N_{dss} + N_{sss}}{N_{qqq}} B(x , z) \nonumber \\
&- & \frac{6 N_{\bar{u}\bar{d}\bar{s}} + 3 N_{\bar{u}\bar{u}\bar{s}} + 3 N_{\bar{d}\bar{d}\bar{s}} + 3 N_{\bar{u}\bar{s}\bar{s}} + 3 N_{\bar{d}\bar{s}\bar{s}} + N_{\bar{s}\bar{s}\bar{s}}}{N_{\bar{q}\bar{q}\bar{q}}} \bar{B}(x,z), \nonumber
\end{eqnarray}
and the strangeness of initial hadrons $S^{(h)} = \sum_\alpha Q_{\alpha,S} N_{\alpha}=-(N_s - N_{\bar{s}})$ because all strange quarks and antiquarks are combined into hadrons after hadronization. Then, we have
\begin{equation}
\begin{split}
C^{(h)}_{BS} = 3 \times \Bigg\{& \frac{24\ B(1,\langle z \rangle)}{(2+\lambda'_s)^3 (1+\langle z \rangle)}  + \frac{24\ \bar{B}(1,\langle z \rangle)}{(2+\lambda_s)^3 (1-\langle z \rangle)}  \\
&+ \frac{12\lambda'_s+6\lambda'^2_s +6\lambda'^3_s}{(2+\lambda'_s)^3} (1+\langle z \rangle)^{a-1} \\
&\times   \frac{\big[(1-\langle z \rangle)^{a-1}(\langle z \rangle^2-2a\langle z \rangle-1)+(1+\langle z \rangle)^{a+1}\big]}{3\big[ (1+\langle z \rangle)^a - (1-\langle z \rangle)^a \big]^2} \\
&+ \frac{12\lambda_s+6\lambda^2_s +6\lambda^3_s}{(2+\lambda_s)^3}  (1-\langle z \rangle)^{a-1}\\
&\times   \frac{\big[(1+\langle z \rangle)^{a-1}(\langle z \rangle^2+2a\langle z \rangle-1)+(1-\langle z \rangle)^{a+1}\big]}{3\big[ (1+\langle z \rangle)^a - (1-\langle z \rangle)^a \big]^2}
\Bigg\}. \label{fcbs2}
\end{split}
\end{equation}
for initial hadron system.
This result shows a complex dependence on baryon-meson competition factor $R_{B/M}(0)$ by the factor $a=\frac{1}{3R_{B/M}(0)}+1$, strangeness $\lambda_s$ and the baryon number density of the system by $z$.

\begin{figure}[!htp]
\centering
  \includegraphics[width=\linewidth]{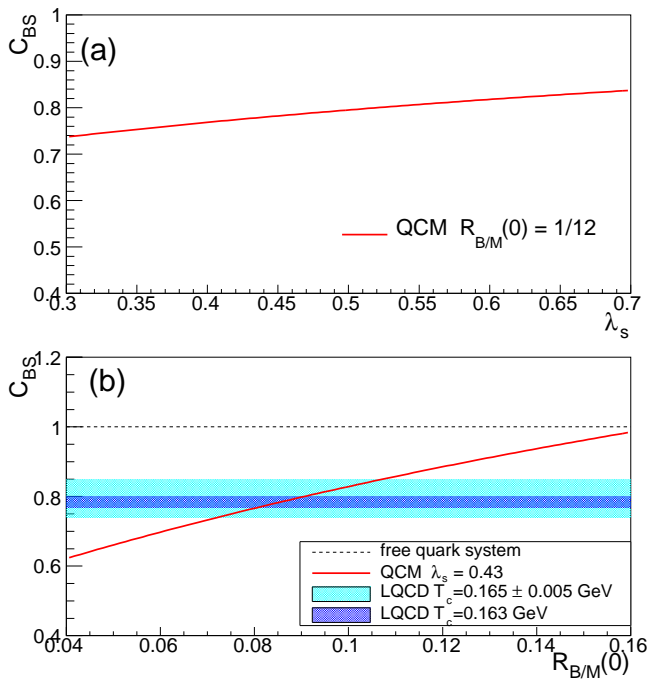}
  \caption{(Color online) The dependence of $C_{BS}$ on strangeness suppression factor $\lambda_s$, (a), and on baryon-meson competition factor $R_{B/M}(0)$, (b). Predictions of Lattice QCD are from Ref.\cite{lqcdCBS}.}
  \label{Cbs2LsRbm}
\end{figure}

In the case of zero baryon number density $z=0$, we have $\lambda'_s = \lambda_s$ and $B(1,0)=\bar{B}(1,0)=1/6a$ and simplify Eq.~(\ref{fcbs2})
\begin{equation}
  C^{(h)}_{BS} = 1+ 8 \frac{ 6 R_{B/M}(0)-1 } {  (2+\lambda_s )^3 (1+ 3 R_{B/M}(0)) }.
\end{equation}
Fig.~\ref{Cbs2LsRbm} (a) shows the dependence of $C^{(h)}_{BS}$ on the strangeness $\lambda_s$, as the $R_{B/M}(0)$ is taken to 1/12. We see that $C^{(h)}_{BS}$ of initial hadrons in the current approach is also insensitive to the strangeness of the system, but the result is about 20\% larger than the previous prediction in Fig.~\ref{Cbsd1LsRbm} (a) in Poisson fluctuation.
Fig.~\ref{Cbs2LsRbm} (b) shows the dependence of $C_{BS}$ on the baryon-meson competition factor $R_{B/M}(0)$. We see that $C^{(h)}_{BS}$ is strongly dependent on the $R_{B/M}(0)$, which is also qualitatively consistent with the result in Fig.~\ref{Cbsd1LsRbm} (b).

The qualitative consistency between Fig.~\ref{Cbs2LsRbm} and Fig.~\ref{Cbsd1LsRbm} can be naturally understood via the definition $C_{BS}=-3\langle B S \rangle /\langle S^2 \rangle$. Because both the nominator $\langle BS \rangle$ and denominator $\langle S^2 \rangle$ are dependent on the strangeness mainly by the single-strange hadrons, the strangeness dependence are thus partially offset in their ratio and we observe a weak dependence of $C^{(h)}_{BS}$ on $\lambda_s$. On the other hand, since the nominator $\langle BS \rangle$ receives the larger contributions from multi-strange baryons than the denominator $\langle S^2 \rangle$ which is always dominated by single-strange mesons, we would observe a slightly increase of $C^{(h)}_{BS}$ with the increasing $\lambda_s$.
In addition, the increase of $R_{B/M}(0)$ enhances the production weights of (strange) baryons against mesons and contributes largely to the nominator $\langle BS \rangle$ and we observe a rapid increase of $C^{(h)}_{BS}$ with the increasing $R_{B/M}(0)$.

Since results in current approach hold more generally than those in Sec.~IV under Poisson statistics, we can compare our result (solid line in Fig.~\ref{Cbs2LsRbm} (b)) with the calculation of Lattice QCD \cite{lqcdCBS} (box area) in the phase transition boundary.
With the critical temperature $T_c=0.16-0.17$ GeV, Lattice QCD \cite{lqcdCBS} predicts the $C_{BS}$ of the system is about $0.74-0.85$, which has a striking overlap with our results as $R_{B/M}(0)$ is within $[1/9.5, 1/13.5]$.
In the preferred situation of $R_{B/M}(0) = 1/12$ and $\lambda_s =0.43$, our calculated $C_{BS}$ is 0.78, which best conforms to the Lattice QCD calculations $C_{BS}=0.77-0.80 $ at $T_c=0.163$ GeV within its theoretical uncertainty \cite{lqcdCBS}.
Such a agreement with Lattice QCD calculations suggests that quark combination mechanism reflect some basic dynamics of the hadronization process of the strong-interacting quark system and thus it is indispensable for the hadronization phenomenology.

\begin{figure}[!htp]
\centering
  \includegraphics[width=\linewidth]{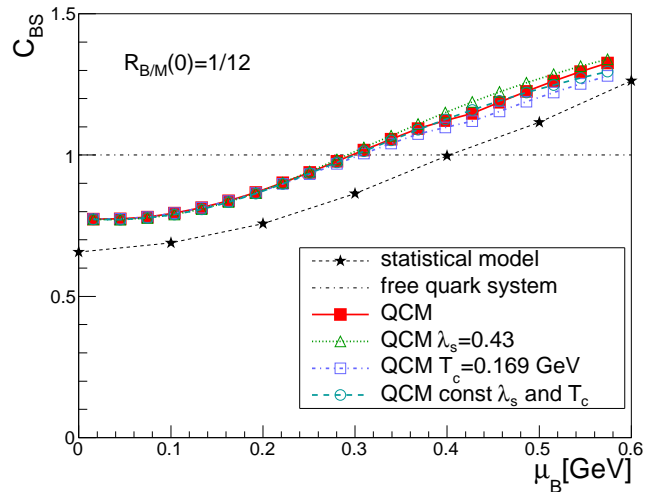}
  \caption{(Color online)  The dependence of $C_{BS}$ on the chemical potential $\mu_{B}$ of the system. The filled squares, up-triangles, open squares and open circles are our results at both varied $\lambda_s$ and $T_c$, at constant $\lambda_s=0.43$, at constant $T_c = 0.166$ GeV, and at both constant $\lambda_s=0.43$ and $T_c=0.166 $ GeV, respectively. They are compared with the prediction of statistic method for a hadron resonance gas \cite{koch05cbs}. }
  \label{cbsmuB2}
\end{figure}

We also predict the dependence of $C_{BS}$ on the baryon number density of the system using the chemical potential $\mu_B$. The results are shown in Fig.~\ref{cbsmuB2}. Here, the transformation of variable $z$ to $\mu_B$ is by Eq.~(\ref{ztomub}) and the $\mu_B$ dependence of $T_c$ and $\lambda_s$ are same as those in Sec.~IV. The results of statistical model from Ref.\cite{koch05cbs} at different $\mu_B$ are also presented as the comparison.
The solid line with filled squares are our results at varied $\mu_B$ and $T_c$.
It is shown that they are about $15\%$ higher than the predictions of statistical model, i.e., dashed line with stars, in the full $\mu_B$ region.
Comparing our results at varied $\mu_B$ and $T_c$ with those at fixed $\mu_B$ and/or $T_c$, we find that decrease of $T_c$ at large chemical potential $\mu_B$ will slightly increase the $C_{BS}$ of the system and that the change of system strangeness at different $\mu_B$ slightly influences the $C_{BS}$ of the system.
These effects of $\mu_B$ and $\lambda_s$ on the $C_{BS}$ of the system are similar to those calculated in Sec.~IV. In addition, we see that results of small $\mu_B$ region calculated here are higher than those in Fig.~\ref{cbsmub1} in Sec.~IV but the results at large $\mu_B$ are close to the previous ones. We also note that another difference between the current calculations of $C_{BS}$ and the previous ones in Fig.~\ref{cbsmub1} is the curvature of $C_{BS}$ as the function of $\mu_B$. The $C_{BS}$ in Fig.~\ref{cbsmub1} is the concave function at intermediate and large $\mu_B$ while the current $C_{BS}$ in Fig.~\ref{cbsmuB2} is almost linear function at intermediate and large $\mu_B$.
By noticing that the predictions of statistical model in this region is a slightly convex function, these different predictions may be identified in future by calculating the higher-order susceptibilities for the conservative charges of the system.
The other difference between QCM prediction and statistical model is the position that the $C_{BS}$ of hadrons reaches one. QCM predicts the position at $\mu_{B} \approx 0.3$ GeV corresponding to the AA collisions at collisional energy about 12 GeV while statistical model predicts the position at $\mu_{B} \approx 0.4$ GeV corresponding to the collisional energy about 9 GeV. The Beam Energy Scan programme of STAR collaboration can distinguish these two different predictions.

\section{Summary and discussion}

Hadronization describes the process of the formation of hadrons out of quarks and/or gluons. In the meanwhile, the correlation properties between conservative charges of the system change also due to the transformation of basic degrees of freedom of the system.
Phenomenological models mechanisms of the hadronization should reproduce these changes of charge correlation properties during hadronization process for the system.
In this paper, we have studied the change of the baryon-strangeness correlation caused by hadronization in the quark combination mechanism. We found that this correlation is a good quantity of studying the hadronization because of its sensitivity on the dynamics of production competition between baryons and mesons.
We calculated the correlation coefficient $C_{BS} = -3(\langle B S \rangle -\langle B\rangle \langle S\rangle)/\langle S^2 \rangle$ of initial hadrons produced from the deconfined free quark system with $C^{(q)}_{BS}=1$.
The calculated $C^{(h)}_{BS}$ of initial hadrons under Poisson fluctuation is about 0.65 at zero baryon chemical potential, which is consistent with the prediction of statistical model for the hadron resonance gas.
Beyond Poisson statistics, we calculated $C^{(h)}_{BS}$ of initial hadrons by $C^{(h)}_{BS}=-3 (\partial B^{(h)}_s /\partial S^{(h)})_L$.
The resulting $C^{(h)}_{BS}$ at zero baryon chemical potential is about 0.78, comparable to the calculations of Lattice QCD in the phase boundary at $T_c =163$ MeV. This suggests that the quark combination, as a phenomenological microscopic mechanism, is able to describe the change of conservative charge correlations at hadronization, revealing certain basic dynamics of the realistic hadronization process.  We also predicted the correlation coefficient of hadrons at different baryon chemical potentials and compared them with the existed calculations of statistical method. These predictions are expected to be tested by the Beam Energy Scan experiment of STAR Collaboration at RHIC and/or by the future Lattice QCD calculations at non-zero chemical potentials.

Finally we would like to comment the measurement of $C_{BS}$ directly by the definition Eq.~(\ref{def_cbs})  either in realistic experiments or in model simulations using an event generator, which already has lots of discussions in literatures. To measure the $C_{BS}$ in general we usually have to select a finite accept window, e.g.~mid-rapidity region $|y|<0.5$.  Selecting a small window size will increase the Poisson statistical fluctuations while selecting a wide window size will inevitably involve the global conservation effect of baryon number and in particular of the strangeness.  Amending such a finite window effect is usually complex.  These considerations just prompt us to calculate the $C_{BS}$ of the system in Sec.~V by the response of baryon number of the system with respect to the change of the system strangeness. This method is less relevant to the accept window size and thus maybe more closely reflect the intrinsic baryon-strangeness correlation of the hadron system.

\section*{Acknowledgments }
The work is supported in part by the National Natural Science Foundation of China under grant 11175104 and 11305076, and by the Natural Science Foundation of Shandong Province, China under grant  ZR2012AM001.

\end{document}